\documentclass[aps,amssymb,12pt,floatfix]{revtex4}
\usepackage{subfigure}
\usepackage{graphicx}
\usepackage{rotating}

\begin{document}

\title{Influence of surface interactions on folding and forced unbinding of semiflexible chains}

\author{V. Barsegov$^1$ and D. Thirumalai$^{1,2}$}
\thanks{Corresponding author phone: 301-405-4803; fax: 301-314-9404; thirum@glue.umd.edu}
\affiliation{$^1$Biophysics Program, Institute for Physical Science and Technology\\
$^2$Department of Chemistry and Biochemistry, University of Maryland, 
College Park, MD 20742}

%\pacs{}
\date{\today}

\begin{abstract}

We investigate the folding and forced-unbinding transitions of adsorbed semiflexible polymer 
chains using theory and simulations. These processes describe, at an elementary level, a 
number of biologically relevant phenomena that include adhesive interactions between proteins 
and tethering of receptors to cell walls. The binding interface is modeled as a solid 
surface and the worm-like chain (WLC) is used for the semiflexible chain (SC). Using Langevin 
simulations, in the overdamped limit, we examine the ordering kinetics of racquet-like 
and toroidal structures in the presence of attractive interaction between the surface and the 
polymer chain. For a range of interactions, temperature, and the persistence length $l_p$ we obtained
the monomer density distribution $n(x)$ ($x$ is the perpendicular distance of a tagged chain end
from the surface) for all the relevant morphologies. There is a single peak in $n(x)$ inside 
the range of attractive forces $b$ for chains in the extended conformations while in racquet and 
toroidal structures there is an additional peak at $x\approx b$. The simulated results for $n(x)$ are 
in good agreement with theory. The formation of toroids on the surface appears to be a first order 
transition as evidenced by the bimodal distribution in $n(x)$. The theoretical result underestimates 
the simulated $n(x)$ for $x\ll b$ and follows $n(x)$ closely for $x\ge b$; the density calculated
exactly agrees well with $n(x)$ in the range $x\ll b$. Chain-surface interaction is probed by 
subjecting the surface structures to a pulling force $f$. The average extension $\langle x(f)\rangle$ 
as a function of $f$ exhibit sigmoidal profile with sharp all-or-none transition at the unfolding 
force threshold $f$$=$$f_c$ which increases for more structured states. Simulated 
$\langle x (f)\rangle$ compare well with the theoretical predictions. The critical force $f_c$ 
is a function of $l_s/l_c$ for a fixed temperature, where $l_c$ and $l_s$ are the length scales 
that express the strength of the intramolecular and SC-surface attraction, respectively. For a 
fixed $l_s$, $f_c$ increases as $l_p$ decreases.  

\end{abstract}
\maketitle

\section{Introduction}

Interactions between biomolecules and surfaces are important in a
number of biological phenomena. Binding and unbinding of proteins from macromolecular complexes
are involved in the regulation of biological functions \cite{1,2,3}. Adsorption of fibrinogen 
influences the adhesion of leukocytes, microphages or platelets. In addition, interaction 
between proteins, DNA and RNA are mediated by biological membranes \cite{4,5,6}. In the 
crowded cellular conditions, protein-protein and  DNA-protein interactions take place in 
confined geometries in which surface interactions are vital. For instance, interaction between 
P-selectin receptors and their specific ligands is mediated by a flat and shallow binding interface 
\cite{7,8}. Besides these situations, which are obviously relevant in biology, there are
a number of situations in polymer science where interactions with surfaces are important 
\cite{9,10,11,12}. These include nanolubrication that involve interaction between surfaces 
that are mediated by polymers. Design of nanoscale materials and biologically inspired 
self-assembling systems also requires an understanding of how heteropolymers and biomolecules 
interact with surfaces. Recent advances in atomic force microscopy \cite{7,13,14} has allowed a 
direct probe of the energetics of interaction between adsorbed proteins with other biomolecules  
\cite{1,15,16,17,17new,18}. The potential applications of polymer-surface models to a number of 
problems has prompted us to develop a theoretical approach which can be used in conjunction with AFM 
experiments to decipher biomolecule-surface interactions.

There have been numerous studies of adsorption of flexible polymers adsorbed on solid surfaces which
find applications in many aspects of colloidal and interface science \cite{10,11,12,19}. However, 
many biomolecules, including DNA, RNA, and proteins, are better described using worm-like 
chain (WLC) models \cite{1,20,21}. Thus, it is important to provide a theoretical 
description of the interactions between semiflexible chains \cite{21,22,22new} and interfaces. 
The purpose of this paper is to address the following specific questions: (i) It is known that DNA, 
a semiflexible polymer, undergoes a coil-globule transition in the presence of osmolytes or 
multivalent cations \cite{22,23}. Simulations of semiflexible chains in poor solvents \cite{19,24} 
have been used to understand the kinetics and pathways of transitions from extended conformations 
to collapsed toroidal structures. The coil-globule transition in stiff chains in the bulk occurs 
through a series of metastable racquet structures \cite{19,24}. How does the 
interaction with the surface alter the morphology and kinetics of such transitions? This question 
is relevant even for DNA collapse in cells where the DNA compaction takes place in the presence of 
interactions with their large biomolecules in restricted spaces; (ii) AFM experiments are likely 
to provide the most direct data for the strength of interaction between semiflexible biomolecules. 
In these experiments one of the molecules of interest is anchored onto the surface while force 
is applied to the end of the other. The unbinding force can be calculated from the 
force-extension profiles. These experiments raise the question, namely, what are the adhesive 
forces between semiflexible polymer and a surface? We address these questions using theory 
\cite{12,25,26} and simulations for a WLC model interacting with a solid surface.

In the absence of the surface the morphologies of semiflexible chain (SC) is determined by 
thermal fluctuations and an interplay of the chain persistence length $l_p$ and intramolecular
condensation length $l_c$$=$$\sqrt{l_p k_B T/u_m}$ where $T$ is the temperature and $u_m$ is the 
effective intramolecular attractive energy per unit length \cite{19}. In the presence of a surface 
another length scale $l_s$$=$$\sqrt{l_p k_B T/u_{ads}}$, where $u_{ads}$ is the attractive SC-surface 
interaction energy per unit length, plays an essential role in the determination of the 
structures. The interplay of $l_p$, $l_c$, and $l_s$ will determine the morphology of the 
surface-induced structures. It also follows that the response to applied force measured in terms 
of force-extension profiles will depend on  $l_p$, $l_c$, and $l_s$. In this paper we explore a 
range of values of $k_B T$ and $l_s$ to predict the force-extension curves for semiflexible
chains in poor solvent.

\section{Theory}

Consider a semiflexible chain interacting with a flat surface with the SC-surface potential being
$U_{ads}$. Force ${\bf f}$$=$$f$$\cdot$${\bf n}_x$ is applied to one end of the chain (Fig. 1). 
Equilibrium chain configuration is described using the conditional probability 
$G({\bf{x}}_N,{\bf{x}}_1;{\bf f})$ of finding the tagged $N^{th}$ monomer at ${\bf{x}}_N$ given 
that the monomer ${\bf{x}}_1$ is anchored at the surface, where ${\bf{x}}$$=$$({\bf{r}},{\bf{n}})$ 
includes position vector ${\bf r}$$=$$(x,y,z)$ and orientation vector ${\bf n}$, respectively. 
Due to axial symmetry, the free end orientation is specified by the angle $\theta$ between its 
tangent vector and the $x$-axis and distance from the surface $x$ (Fig. 1), and the conditional 
probability $G(x_N,\theta_N;x_1,\theta_1;f)$ can be used instead of $G({\bf{x}}_N,{\bf{x}}_1;{\bf f})$. 
In the limit $L$$\to$$\infty$, $G(x_N,\theta_N;x_1,\theta_1;f)$ is dominated by the ground state 
$\psi_0$, so that
\begin{equation}\label{1.1}
G(x_N,\theta_N;x_1,\theta_1;f)\approx \psi_0 (x_1,\theta_1;f)\psi_0^{\dagger} (x_N,\theta_N;f) 
\exp{[-\beta N \epsilon_0 ]}
\end{equation}
$\epsilon_0$$=$$E_0$$/$$N$ is the equilibrium free energy per monomer and $\beta$$=$$1$$/$
$k_B T$. If the SC is modeled as a worm-like chain (WLC), then $\psi_0$ satisfies \cite{21,25,26},
\begin{equation}\label{1.3}
-2l_p \gamma {{\partial \psi_0}\over {\partial x}} + 
(1-\gamma^2){{\partial^2 \psi_0}\over {\partial \gamma^2}}-
2\gamma {{\partial \psi_0}\over {\partial \gamma}} + \beta l_p \gamma f\psi_0 = 
\beta (u_{ads}(x)-\epsilon_0)\psi_0
\end{equation}
where $u_{ads}=U_{ads}/N$ is the adsorption potential per monomer and $\gamma=\cos{[\theta]}$
(see Fig. 1). To mimic the Lennard-Jones chain-surface attractive interaction used in the
Langevin simulations (see Section III), we employ a piece-wise continuous potential, i.e. 
$u_{ads}$$=$$\infty$ for $x$$<$$0$, $u_{ads}$$=$$-\Delta$ for $0$$\le$$x$$\le$$b$ and 
$u_{ads}$$=$$0$ for $x$$>$$b$. The monomer density of the adsorbed structures in the absence of force,
\begin{equation}\label{1.3a}
n(x)=\int d\theta  \psi_0^2(x,\theta)
\end{equation}
normalized as $\int dx$$n(x)$$=$$N$, is calculated by solving Eq. (\ref{1.3}) without the last 
term $\beta$$l_p$$\gamma$$f$$\psi_0$. 

The perturbative solution of Eq. (\ref{1.3}) in the absence of the $\beta$$l_p$$\gamma$$f$ term,
due to Kuznetsov and Sung \cite{25}, to the first order in correlation length parameter 
$\eta$$=$$(4l_p^2/{\psi}_0)$$\left| d^2 {\psi}_0/dx^2\right|$ is outlined in Appendix A. The solution 
is
\begin{eqnarray}\label{2.13}
{\psi}_0 (x) & = & h(b-x)\large(C_1\sin{\left({{x\sqrt{m_1}}\over {2l_p}}\right)}+
C_2\sin{\left({{x\sqrt{m_2}}\over {2l_p}}\right)}\\\nonumber
& + & C_3\cos{\left({{x\sqrt{m_1}}\over {2l_p}}\right)}-
C_3 \cos{\left({{x\sqrt{m_2}}\over {2l_p}}\right)}\big), \\\nonumber
{\psi}_0 (x) & = & C_0 h(x-b)\exp{\left(-{{x\sqrt{k}}\over {2l_p}}\right)}
\end{eqnarray}
where $h(x)$ is the Heaviside function, and $C_0$, $C_1$, $C_2$ and $C_3$ are constant coefficients; 
$m_1$, $m_2$ and $k$ are given by
\begin{eqnarray}\label{2.14}
m_{1,2} & = & {{15}\over {8}}(2+\phi_{in})(6+\phi_{in})\left( 1\mp 
\sqrt{1+{{16}\over {5}}{{\phi_{in}}\over {6+\phi_{in}}}}\right) \\\nonumber
k & = & {{15}\over {8}}(2+\phi_{out})(6+\phi_{out})\left( -1+ 
\sqrt{1+{{16}\over {5}}{{\phi_{out}}\over {6+\phi_{out}}}}\right)
\end{eqnarray}
and $\phi_{in}$$=$$\beta$$(u_{ads}(x)$$-$$\epsilon_0)$$<$$0$, $\phi_{out}$$=$$-\epsilon_0$$>$$0$.
By using two continuity requirements (\ref{2.11}) and the normalization, we can obtain, respectively,
$\epsilon_0$ and $C_0$ and one of $C_1$, $C_2$ or $C_3$. However, the two free constants are to be 
chosen such that a minimum of $\epsilon_0$ is obtained. The minimal free energy corresponding to the 
ground state for $x\le b$ is attained for (i) $C_1$$\neq$$0$, $C_2$$=$$C_3$$=$$0$ (i.e. the state with 
$m$$=$$m_1$) and (ii) $C_2$$\neq$$0$, $C_1$$=$$C_3$$=$$0$ ($m$$=$$m_2$).  

Perturbative solution of Eq. (\ref{1.3}) ignores variation of $\psi_0$ on $\theta$. Indeed,
when $x$$\gg$$l_p$, $\psi$ becomes nearly isotropic, $\psi$$(x,\theta)$$=$$\psi (x)$. However, 
when $x$$\sim$$b$$\ll$$l_p$, $\psi_0$ should strongly depend on the angle $\Theta=\pi/2+\theta$ 
between free end of the chain and the surface (Fig. 1). In this range 
$-\gamma$$\partial$$/\partial$$x$$\to$$\Theta$$\partial$$/\partial$$x$, $(1-\gamma^2)$
$\partial^2$$/\partial$$\gamma^2$$\to$$\partial^2$$/\partial$$\Theta^2 $, $\gamma$$\partial$
$/\partial$$\gamma$$\to$$0$, $\gamma$$f$$\to$$-$$\Theta$$f$ and Eq. (\ref{1.3}) simplifies, i.e.
\begin{equation}\label{3.1}
\Theta {{\partial \psi_0}\over {\partial x}} + 
{{1}\over {2l_p}}{{\partial^2 \psi_0}\over {\partial \Theta^2}} - {{1}\over {2}}\Theta f \psi_0
={{\beta}\over {2l_p}}(u_{ads}(x)-\epsilon_0)\psi_0.
\end{equation}
The methodology for solving Eq. (\ref{3.1}) has been presented by Semenov in Ref.~\cite{26} 
and is outlined in Appendix B. The general solution for $x$$>$$b$ ($u_{ads}$$=$$0$) is
\begin{equation}\label{3.7}
{\psi}_0 (x,\Theta)=\sum_{n=0,1}C_n x^{1/6-n} \Psi\left( n-{{1}\over {6}},{{2}\over {3}},
{{2l_p \Theta^3 }\over {9x}}\right)
\end{equation}
where $C_0$, $C_1$ are constants and the confluent hypergeometric function $\Psi(\chi,\omega,z)$ 
is $\Psi(\chi,\omega,z)$$\equiv$${{1}\over {\Gamma (\chi)}}$$\int_0^{\infty}$$d\tau$
$\tau^{\chi-1}$$(1+\tau)^{\omega-\chi-1}$$e^{-\tau z}$ where 
$\Gamma (\chi)$$\equiv$$\int_0^{\infty}$$d\tau$$\tau^{\chi-1}$$\exp{[-\tau]}$ is Gamma
function \cite{27}. To describe the chain in the range $x\le b$, we assume that ${\psi}_0$ is 
of the form (\ref{3.7}) and $C_0$ and $C_1$ depend on $x$. Substituting Eq. (\ref{3.7}) into 
Eq. (\ref{3.1}), we obtain:
\begin{eqnarray}\label{3.8}
{{dC_0}\over {dx}} & = & \phi(x) \left( {{2l_p}\over {x}} \right)^{1/3}
\left( F_{00}C_0 + F_{01}x^{-1}C_1\right)\\\nonumber
{{dC_1}\over {dx}} & = & \phi(x) \left( {{2l_p}\over {x}} \right)^{1/3}
\left( F_{10}x C_0 + F_{11}C_1\right)
\end{eqnarray}
where $F_{nm}$$\equiv$$(g_m/\kappa ,f_n)$$/(f_n,f_n)$, $n,m=0,1$, and 
$(g_n,g_m)$$\equiv$$\int_{-\infty}^{\infty}$$d\kappa$$\kappa$$e^{-\kappa^3/9}$$g_n(\kappa)$
$g_m(\kappa)$ (see Eq. (\ref{3.4})). We solve Eqs. (\ref{3.8}) subject to the condition 
$\phi(x)$$=$$\phi_{in}$ for $x$$\le$$b$ and $\phi(x)$$=$$\phi_{out}$ for $x$$>$$b$. From the 
solutions of Eqs. (\ref{3.8}) in Appendix B we obtain:
\begin{eqnarray}\label{3.10}
C_1^{in}(x) & = & c_2 e^{ {{3}\over {2}}\sqrt{-D} x^{{{2}\over {3}}} }\left(-3\sqrt{-D} 
x^{{{2}\over {3}}} \right)^{{{3}\over {2}}} \Phi \left(\rho+{{3}\over{2}},{{5}\over{2}},
-3\sqrt{-D} x^{{{2}\over {3}}}\right) h(b-x) \\\nonumber
C_1^{out}(x) & = & c_1 e^{-{{3}\over {2}}\sqrt{-D} x^{{{2}\over {3}}} }\Phi 
\left( \rho,-{{1}\over{2}}, 3\sqrt{-D} x^{{{2}\over {3}}}\right) h(x-b) 
\end{eqnarray}
where $\sqrt{-D}=\phi_{out} (2l_p)^{{{1}\over {3}}}\sqrt{F_{01}F_{10}-F_{11}F_{00}}$ 
and $h(x)$ is Heaviside step function. In Eq. (\ref{3.10}) the Kummer function $\Phi (k,l,x)$ 
is defined by $\Phi (k,l,x)$$\equiv$$1$$+$$\sum_{m=1}^{\infty}$${{(k)_m}\over {(l)_m}}$
${{x^m}\over {m!}}$ with $(k)_0$$=$$1$, $(k)_1$$=$$k$ and $(k)_m$$=$$k(k+1)$$\ldots$$(k+m-1)$ 
\cite{27}. We get $C_0(x)$ by substituting Eqs. (\ref{3.10}) for $C_1^{in}$ and $C_1^{out}$ into 
the second Eq. (\ref{3.8}) and $\psi_0(x,\Theta ;f)$ can now be obtained by using Eq. (\ref{3.7})).

In the presence of pulling force, $\psi_0$ is nearly isotropic, i.e. $\psi_0(x,\theta;f)$$\approx$
$\psi_0(x;f)$. This allows us to analyze force-extension profiles by employing perturbative 
treatment outlined above. Solution to Eq. (\ref{1.3}) is given by
\begin{equation}\label{2.1}
\psi_0(x;f)={\psi}_0(x)e^{{{1}\over {2}}\beta f x}
\end{equation}
where ${\psi}_0(x)$ is given by Eqs. (\ref{2.13}). The average extension as a function of applied 
force can be computed using 
\begin{equation}\label{1.5}
\langle x (f)\rangle \equiv 
{{1}\over {\beta}}{{1}\over {Z(f)}}{{d}\over {df}}Z(f)
\end{equation}
where the partition function $Z(f)$ is 
$Z(f)$$=$$\int d\theta_N$$\int d\theta_1$$\int dx_N$$\int dx_1$$G(x_N,\theta_N;x_1,\theta_1;f)$.

The perturbation theory is strictly valid only when the condensation length $l_c$$\gg$$l_p$. In
practice we find that the first order perturbation theory gives results that are in very good
agreement with simulations even when $l_c$$\sim$$l_p$. Kuznetsov and Sung also discovered that the
perturbation theory is remarkably successful outside the regime of applicability \cite{25}.

\section{Langevin dynamics simulations}

We model a semiflexible chain (SC) by $N$$=$$100$ connected beads of bond length 
$a$ and the contour length $L$$=$$100$$a$. In the absence of $U_{ads}$ and $f$$=$$0$, we assume 
that the dynamics is governed by the overdamped Langevin equation:
\begin{equation}\label{0.1}
\xi {{d}\over {dt}}{\bf x}_j=-{{\partial U}\over {\partial {\bf x}_j}}+{\bf g}_j (t)
\end{equation}
where $\xi$ is the friction coefficient, $U$$=$$U_{chain}$$=$$U_{bond}$$+$$U_{bend}$$+$$U_{LJ}$ 
is chain internal energy due to bond potential $U_{bond}$, bend potential $U_{bend}$ 
and interbead interaction potential $U_{LJ}$ (hydrodynamic interactions are ignored).
The random force ${\bf g}_j (t)$ obeys Gaussian statistics,
\begin{equation}\label{0.2}
\langle {\bf g}_j (t)\rangle = 0, \quad  
\langle {\bf g}_i (t){\bf g}_j (t')\rangle = 6 k_B T \xi \delta_{ij} \delta(t-t')
\end{equation}
We solve Eq. (\ref{0.1}) for each ${\bf x}_j$ with unit tangent vector 
${\bf u}_j$$=$$({\bf x}_{j+1}$$-$${\bf x}_j)/a$, where $j$$=$$1,2,\ldots ,N$. The stretching
potential $U_{bond}$ is 
\begin{equation}\label{0.3}
U_{bond} = {{A}\over {2\sigma^2}}\sum_{j=1}^{N-1}(|{\bf x}_j - 
{\bf x}_{j+1}|^2-\sigma)^2,
\end{equation}
where $A$ and $\sigma$ are constants, and 
\begin{equation}\label{0.4}
U_{bend} = {{S}\over {2}}\sum_{j=1}^{N-1}\left( 1+\cos{[\varphi_{j,j+1} ]} \right)^2
\end{equation}
where the constant $S$ is a measure of chain stiffnes, and $\cos{[\varphi_{j,j+1} ]}$$=$
$({\bf x}_{j+1}$$-$${\bf x}_j)$$({\bf x}_{j-1}$$-$${\bf x}_j)$$/$$\sigma^2$ is the bend
angle. The interaction between beads is given by the 12-6 Lennard-Jones potential,
\begin{equation}\label{0.6}
U_{LJ}=B \sum_{i<j}\left[  \left( {{\sigma }\over {\Delta {\bf x}_{ij}}} 
\right)^{12} - 2 \left( {{\sigma }\over {\Delta {\bf x}_{ij}}}\right)^6 \right]
\end{equation}
where $\Delta {\bf x}_{ij}$ is the distance between beads $i$ and $j$, and $B$ 
is the magnitude of interaction. $U_{LJ}$ is an effective interaction that accounts for excluded 
volume interactions and counterion induced attraction which in DNA is due to screening of the
charges. The persistence length of the chain $l_p$ can be roughly estimated by using 
$l_p$$=$$a$$/$$(1-\cos{[\langle \varphi_{j,j+1}\rangle ]})$ where 
$\langle$$\varphi_{j,j+1}$$\rangle$$=$$(N-1)^{-1}$$\sum_{j=1}^{N-1}$$\varphi_{j,j+1}$ 
is the average angle between adjacent beeds. 

Similar models have been used in previous studies to probe the
chain collapse in poor solvents \cite{19,24}. In the presence of the adsorbing surface the 
motion of $j$-th bead is governed by Eq. (\ref{0.1}) with $U$$=$$U_{chain}$$+$$U_{ads}$, where 
$U_{ads}$ is the surface-SC potential,
\begin{equation}\label{0.8}
U_{ads}=\Delta \sum_{i}\left[  \left( {{b }\over {\Delta {\bf x}_{i}}} 
\right)^{12} - 2 \left( {{b }\over {\Delta {\bf x}_{i}}}\right)^6 \right].
\end{equation}
In Eq. (\ref{0.8}) $\Delta {\bf x}_{i}$ is the bead-surface distance and $\Delta$ and $b$ 
are, respectively, the depth and range of the attractive forces. We set $B=1.0$,
$\sigma$$=$$a$$=$$1$ and $b$$=$$3$$a$, and use $A$$=$$400$$B$, $S$$=$$30$$B$, $60$$B$, $120$$B$ 
and $\Delta$$=$$1.5$$B$, $2.0$$B$, $2.5$$B$. This makes $U_{LJ}$, $U_{bond}$, $U_{bend}$ and 
$U_{ads}$ to scale in units of $\epsilon_h$$=$$k_B T$ and $\epsilon_l$$=$$\sigma$ is the 
unit length. The choice $A$$=$$400$$B$ allows for $5$ percent of thermal fluctuations in 
the bond distance and permits us to run simulations with longer time steps without affecting bond 
relaxation time. The unit of time is $\tau$$=$$\xi \sigma^2/\epsilon_h$, where 
$\xi$$=44.0$ is the friction coefficient of the chain in water at $T=300 K$. The system of 
equations (\ref{0.1}) is integrated with a step size $\delta t$$=$$2$$\times 10^{-2}$$\tau$ and 
the total time is $t$$=$$N_{tot}$$\delta t$ where $N_{tot}$ is the number of integration steps.
We express time either in units of $\tau$ or in terms of $N_{tot}$.

\section{Results}
\subsection{Surface-induced structural transitions:}

It is known that in the absence of the surface SC undergoes a collapse transition
when the solvent is poor i.e., when the attractive monomer-monomer interactions dominate 
(Eq. (\ref{0.6})) so that $l_c$$>$$l_p$. The collapse is a result of a competition between 
intramolecular attraction and bending energy due to chain stiffness. Unlike in flexible polymers, 
the low energy collapsed conformation is a torus which maximizes intramolecular contacts and 
minimizes the bending penalty. Before simulating the force-extension curves of adsorbed SC it is 
necessary to characterize the structures that are obtained when interacting with the surface. 

To simulate the low free energy structures that result
in the presence of the surface, we first thermalized an extended chain at high temperature 
$k_B T$$=$$3.0$ for $N_{tot}$$=$$1$$\times$$10^6$ steps. By gradually decreasing the temperature 
bulk structures were thermalized for $(1-10)$$\times$$10^7$ steps and used in adsorption experiments. 
Interactions with the attractive surface was switched on at distance $\Delta$$x$$=$$2b$ away from 
the bead with shortest $x$ and the SC was adsorbed onto the surface one bead at a time. The structures 
were allowed to relax for $\sim$$1$$-$$20$$\times$$10^6$ steps depending on $k_B T$, $S$ and 
$\Delta$. Progress of adsorption was monitored by analyzing time traces of $U_{LJ}$, $U_{ads}$, 
$U$, and the radius of gyration $R_g$ of the SC. We generated $500$ adsorbed structures at 
$k_B T$$=$$1.0$, $1.25$ and $1.5$ for $S$$=$$30$$B$, $60$$B$, $120$$B$ and $\Delta$$=$$1.5$$B$, 
$2.0$$B$ and $2.5$$B$. 

Typical structures are presented in Fig. 2. Geometry of the SC adsorbed onto the surface
ranges from partially or fully extended configuration with $l_p/a\approx 18$ to partially 
structured one-, two- and three-racquet states with $l_p/a$$\approx$$16.0$, $15.5$ and $15.0$, 
respectively, to fully ordered toroidal states with $l_p/a\approx 13.5$. Similar structures have 
been observed in recent studies of collapse of semiflexible chains in the bulk \cite{19,24}. For
the interaction parameters used in our simulations $l_s/l_c$$\sim$$o(1)$. Thus, the attractive
SC-surface interaction facilitates adsorption of the SC without significantly altering its
morphology compared to the bulk case. For $l_s$$\gg$$l_c$ the lowest free energy structures are
extended.

To compare the kinetics of structure formation on the surface and in the bulk we also simulated 
collapsed structures in the absence of the adsorbing surface. By analyzing the temporal profiles of 
$R_g$, $U_{LJ}$ and $l_p$, we found that on average, chains attain structured configurations on a 
faster timescale when adsorbed on the surface. The search for the ground (toroidal) state is more 
efficient when the chain is constrained to evolve on the two-dimensional surface where the SC 
quickly minimizes its free energy in reduced $d$$=$$2$-space by sliding surface motion (lateral 
diffusion).

\subsection{Kinetics of surface-induced ordering:}

Typically, surface-induced ordered structures form by a two step process $B_0$$\to$$S_0$$\to$$S_t$. 
Starting from the bulk state $B_0$, extended surface transient $S_0$ emerges during the fast first 
step with the $B_0$$\to$$S_0$ transition occuring within $N_{tot}$$=$$1-3$$\times$$10^6$. In the 
slower second step $S_0$$\to$$S_t$, extended transient structures explore the free energy landscape 
in search of the toroidal state $S_t$ which occurs in about $3-20$$\times$$10^6$ steps depending on 
$S$, $\Delta$ and temperature. Transition from $S_0$ to $S_t$ is realized via rapid formation of 
either a surface loop or an intermediate toroid-like motif with larger $R_g$ (smaller winding 
number) or through a sequence of longer lived racquet states $S_0$$\to$$S_1$$\to$$\ldots$$\to$$S_t$, 
where $S_n$, $n$$=$$0$,$1$,$2$,$\ldots $ denotes conformations with number of racquets equal to zero
(extended chain) one, two, etc. 

The number of ``metastable'' racquets depends on chain flexibility. We observed configurations with 
$n$$=$$6$ for $S$$=$$30$$B$ and $k_B T$$=$$1.25$.  Simulated profiles of $R_g$, $U_{LJ}$, $U_{ads}$ 
and $U$ indicate that evolution from extended to toroidal states follows several pathways. Four out 
of five simulation runs followed the scheme outlined above. Similar diverse pathways have been 
observed by Noguchi and Yoshikawa \cite{19} who recorded the lifetime of intermediates species for 
about $N_{tot}$$=$$2.0$$\times$$10^5$. Our results indicate that attractive surface forces increase 
the lifetimes of metastable intermediates for stiff chains at low temperature. In few simulation runs 
toroidal structures were not observed during as many as $20$$\times$$10^6$ steps. Hence, attractive 
surface forces facilitate formation of toroidal state primarily when formation of toroid-like 
intermediate motif is involved.

The dynamics of $R_g$, $U_{LJ}$, $U_{ads}$ and $U$ for the structures in Fig. 2 show that 
increasingly more ordered states are also energetically favorable (Fig. 3). $R_g$, $U$ and $U_{LJ}$ 
decrease and $U_{abs}$ increases in the sequence $S_0$$\to$$S_1$$\to$$S_2$$\to$$S_3$$\to$$S_t$. 
$R_g$ fluctuates around larger values for extended states. Variations in $U_{LJ}$, $U_{ads}$ and 
$U$ increasing in the sequence of $S_1$$\to$$S_2$$\to$$S_3$$\to$$S_t$ transitions are due to 
formation of SC-surface contacts. For the structures in Fig. 2, the formation of $S_1$ at 
$N_{tot}$$\approx$$2.0$$\times$$10^6$ is mediated by a surface-loop motif followed by slow sliding 
motion; $S_2$ forms early at $N_{tot}$$\approx$$5.0\times$$10^5$ and remains unchanged (time
dependence of $R_g$ or $U_{LJ}$). The dynamics of $U_{LJ}$, $U_{ads}$ and $U$ show formation of 
$S_3$ via $S_1$ at $N_{tot}$$\approx$$1.0$$\times$$10^5$ followed by transition $S_1$$\to$$S_3$ 
at $N_{tot}$$\approx$$7.0$$\times$$10^6$. Similarly, traces of same quantities
for $S_t$ point at three step transition, $S_0$$\to$$S_1$$\to$$S_3$$\to$$S_t$ occuring respectively 
at $N_{tot}$$\approx$$5.0$$\times$$10^5$, $1.5$$\times$$10^6$ and $4.0$$\times$$10^6$, followed by 
chain compaction due sliding motion. 

In agreement with theoretical arguments \cite{26} monomer profiles of stiff chains 
($S$$=$$120$$B$, $l_p/b\gg 1$) are described by the succession of short near-surface loops of  
length $\ll l_p/a$ between chain-surface contact and by the combination of short and long loops of 
the length $\gg l_p/a$ for $S$$=$$30$$B$ and $l_p/b$$\sim$$1$. Decrease in $l_p/b$ and temperature
favors the formation of chain-surface contacts by enabling more beads to be inside the range of 
surface forces. This results in the formation of higher ordered states $S_4$, $S_5$, $S_6$ and $S_t$. 
In contrast, at higher temperatures and increased $\Delta$ and $S$, surface structures with increased 
conformational free energy become unstable and unfold into extended configurations (data are 
not shown). We quantified the geometry of surface structures (Fig. 2) by binning bead-surface 
distances $x_j$ into the density histogram $n(x)$. The monomer density profiles for 
$\Delta$$=$$1.5$$B$ and $k_b T$$=$$1.0$ are compared in Fig. 4 for $S$$=$$70$$B$ (left) and 
$S$$=$$50$$B$ (right panels). Transition from less structured to more structured states is 
accompanied by an increased ratio of the number of bead-bead to bead-surface contacts. The density 
distribution $n(x)$ is single-peaked at $x$$=$$b/2$ and decays to zero as $x$$\to$$b$ for extended 
states and increases its density at $x$$\approx$$b$ in the sequence 
$S_1$$\to$$S_2$$\to$$S_3$$\to$$S_t$. 

\subsection{Forced unfolding of surface adsorbed structures:}

To unfold the surface-ordered structures, these structures were initially allowed to thermalize at 
$k_B T$$=$$1.0$ for $N_{tot}$$=$$2\times 10^6$. We then ancored the $C$-terminus of the chain at 
the surface and pulled its $N$-terminus with constant force $f$ via the harmonic spring with the 
spring constant $k_{sp}$$=$$0.36pNnm^{-1}$ in the direction perpendicular to the surface. Simulation 
runs were terminated after evolution of chain extension $x(N_{tot})$ had reached equilibrium. 
$x(N_{tot})$ of the structures of Fig. 2 are presented in Fig. 5 for $f$$=$$9.75$$pN$, $18.3$$pN$, 
$24.4$$pN$ and $30.5$$pN$. Chain extension reaches saturation plateau in the first $8$$\times$$10^7$ 
steps as the chain restoring force approaches $f$. Not unexpectedly, the unfolding threshold force 
increases as the extent of ordering decreases in the sequence 
$S_0$$\to$$S_1$$\to$$S_2$$\to$$S_3$$\to$$S_t$. At $f$$=$$9.75$$pN$ only 
$S_0$ unfolds in $1.0$$\times$$10^7$ steps. When the force is increased to $f$$=$$18.3$$pN$, $S_0$, 
$S_1$, $S_2$ and $S_3$ unbind from the surface in $3.5$$\times$$10^7$, $3.6$$\times$$10^7$, 
$4.0$$\times$$10^7$ and $6.0$$\times$$10^7$ steps, respectively. At $f$$=$$24.4$$pN$ all structures 
reach the stretched state in $2$$-$$4$$\times$$10^7$ steps. From the dynamical trajectories of
$x$ obtained for $\Delta$$=$$1.5$$B$ and $k_B T$$=$$1.0$, we constructed the average extension 
$\langle$$x$$\rangle$ as a function of $f$. In Fig. 6 we compare $\langle$$x$$\rangle$ vs $f$ traces 
for extended, one-racquet, three-racquet and toroidal structures of Fig. 2 ($S$$=$$120$$B$, top 
panel) and more flexible four-, five- and seven-racquet and toroidal conformations obtained for 
$S$$=$$30$$B$. Unbinding of surface-anchored structures undergo a highly cooperative all-or-none 
transition as the unfolding force threshold $f$$=$$f_c$ is increased from $7.3$$-$$15.8$$pN$ 
($S$$=$$120$$B$) to $15.9$$-$$17.7$$pN$ ($S$$=$$30$$B$) for more compact racquet and toroidal states.

\subsection{Comparison between theory and simulations:}

We analyzed the simulation results for the monomer density and the averaged extension as a 
function of the pulling force by using perturbative treatment 
(see Eqs. (\ref{1.3a})-(\ref{2.14})) in the entire range of $x$$<$$L$. For the proximal limit 
$x$$\ll$$l_p$$<$$L$ we use the exact expressions in Eqs. (\ref{3.7}) and (\ref{3.10}). Density 
distributions and force-extension profiles for the extended conformation were approximated by 
chosing the ground state with $m$$=$$m_1$ ($m_1$$<$$m_2$, see Eq. (\ref{2.14})). The choice 
$m$$=$$m_1$ corresponds to isotropic-like unstructured surface state with no prefered orientation 
of the chain beads. Histograms of structured two-, three-racquet and toroidal conformations were 
analyzed with the choice $m$$=$$m_2$ corresponding to nematic-like ordered states \cite{25}. 
To account for the difference between the shape of attractive potential $U_{ads}$ used in the 
simulations and the theoretical calculation we used, in the actual fit, the rescaled potential 
depth $\Delta_{T}$$=$$r$$\Delta_{sim}$ for the same range $b_{T}$$=$$b_{sim}$$=$$1$, where 
$r$$=$$(b \Delta_T)^{-1}$$\int_0^{\infty}$$dx$$u_{ads}(x)$ is the ratio between volume of Lennard-Jones 
attractive layer and $b$$\Delta_T$ used in theory. The density profiles $n(x)$ for known values of $b$, 
$k_B T$ and $\Delta_{T}$ were fitted to the simulated monomer density histograms and force-extension 
profiles to obtain parameters $\epsilon_0$ (Eq. (\ref{1.1})) and $l_p$. The theoretical results for
the density $n(x)$ and the average extension $\langle x(f)\rangle$ computed from Eqs. (\ref{1.3a}) 
and (\ref{1.5}), respectively, using these parameters are shown in Figs. 4 and 6.

{\it Monomer density distributions:} Although the theoretical results for $n(x)$ slightly underestimate 
the simulated density for structured states for $b$$<$$x/a$$<$$1.5$ and underestimates it for 
$x/a$$>$$1.6$, the agreement between perturbation theory and simulation data is surprisingly good 
in the range of $x/a$$\le$$b$ (Fig. 4). The agreement between theory and simulations improves for 
more structured racquet and toroidal conformations. In particular, the theoretical profiles capture 
the positions of density peaks both inside the layer at $x/a$$\approx$$0.5$ and at the 
boundary $x/a$$\approx$$b$. Although there is some residual density at large $x/a$ due to thermal 
fluctuations of chain ends, especially for less structured extended and racquet configurations, the 
ground state dominance approximation is clearly valid. The theoretically estimated conformational 
free energy per monomer and persistence length for structures $S_0$$\to$$S_2$$\to$$S_3$$\to$$S_t$ 
of Fig. 4 decrease respectively as 
$\epsilon_0/k_B T$$\approx$$-$$0.21$$\to$$-$$0.23$$\to$$-$$0.24$$\to$$-$$0.25$ 
and $l_p/a$$\approx$$11.7$$\to$$11.2$$\to$$10.4$$\to$$10.2$ (for $k_B T$$=$$1.0$, left panels), and 
$\epsilon_0/k_B T$$\approx$$-$$0.23$$\to$$-$$0.25$$\to$$-$$0.26$$\to$$-$$0.27$ 
and $l_p/a$$\approx$$11.3$$\to$$11.0$$\to$$10.2$$\to$$10.1$ (for $k_B T$$=$$1.25$, right panels).
Not surprisingly, both $\epsilon_0$ and $l_p$ decrease for the same structures as $k_B T$ is 
increased because of enhanced chain flexibility. In the proximal region, the exact calculation of
$n(s)$ for $0$$\le$$x/a$$\le$$0.5$ for the same structure sequence shows a better agreement with 
the simulated results. The fit parameters are 
$\epsilon_0/k_B T$$\approx$$-$$0.2$$\to$$-$$0.24$$\to$$-$$0.25$$\to$$-$$0.28$ 
and $l_p/a$$\approx$$12.3$$\to$$11.8$$\to$$11.0$$\to$$10.8$ for $k_B T$$=$$1.0$, and 
$\epsilon_0/k_B T$$\approx$$-$$0.19$$\to$$-$$0.22$$\to$$-$$0.24$$\to$$-$$0.26$ 
and $l_p/a$$\approx$$12.0$$\to$$11.6$$\to$$10.8$$\to$$10.6$ for $k_B T$$=$$1.25$. 

{\it Force-extension curves:} Apart from small deviations around the unfolding threshold forces 
for all simulated surface structures, the fit of theoretical curves of the average extension vs 
pulling force to simulated data points shows excellent agreement between theory and simulations. 
The theoretical $\langle x(f)\rangle$ curves calculated using perturbation theory follow 
closely the simulated force-extension profiles both for $S$$=$$120$$B$ and $S$$=$$30$$B$ especially 
below ($x/L$$\le$$0.1$) and above ($x/L$$\ge$$0.9$). The unbinding threshold forces increase as 
$7.5pN$$<$$10.5pN$$<$$12.5pN$$<$$16.5pN$ in the sequence $S_0$$\to$$S_1$$\to$$S_3$$\to$$S_t$ 
($S$$=$$120$$B$, top panel in Fig. 6) and as $15pN$$<$$15.5pN$$<$$16.5pN$$<$$17.5pN$ in the sequence
$S_4$$\to$$S_5$$\to$$S_7$$\to$$S_t$ ($S$$=$$30$$B$, bottom panel in Fig. 6). This implies that
more flexible and/or more structured surface chains are harder to unfold. However, ``all-or-none'' 
type of simulated unfolding transition shows sharper growth than predicted by the theory. The 
theoretically estimated conformational free energy per monomer and persistence length for 
structures $S_0$$\to$$S_1$$\to$$S_3$$\to$$S_t$ decrease respectively as
$\epsilon_0/k_B T$$\approx$$-$$0.12$$\to$$-$$0.17$$\to$$-$$0.20$$\to$$-$$0.22$ 
and $l_p/a$$\approx$$15$$\to$$14.5$$\to$$14.25$$\to$$12.1$ (top panel). For the
structures $S_4$$\to$$S_5$$\to$$S_7$$\to$$S_t$, $\epsilon_0$ decreases as 
$\epsilon_0/k_B T$$\approx$$-$$0.134$$\to$$-$$0.136$$\to$$-$$0.141$$\to$$-$$0.148$
and $l_p/a$$\approx$$8.2$$\to$$8.1$$\to$$8.0$$\to$$7.9$ (bottom panel). Here too,
increased chain flexibility decreases $l_p$ and lowers $\epsilon_0$.

\section{Conclusions}

To provide insights into interactions between biomolecules interacting with membranes we have
considered collapse and forced-unbinding of semiflexible chains (SC) in the presence of an adsorbing
surface. The interaction of SC modeled using WLC, which describes well many of the physical properties 
of DNA \cite{18}, RNA \cite{28}, and proteins \cite{29}, with a surface into which the SC can adsorb, 
is studied using theory and simulations. The morphologies of the SC in the presence of an adsorbing 
potential is described in terms of three length scales, namely, $l_p$, $l_s$, and $l_c$. By restricting 
ourselves to $l_c$$\approx$$l_s$ we have studied the effect of interaction with the surface on 
coil-toroidal transition in DNA like chains. The simulations show that the rate of toroid formation 
is impeded compared to the bulk because interaction with the surface stabilizes many metastable 
racquet-like structures (Fig. 1). The simulated equilibrium density profiles show that as the range 
of surface-SC interaction increases and temperature decreases, which leads to a decrease in $l_p/l_s$, 
ordered structures form. The peak of $n(x)$ at $x$$\approx$$b$ (the range of interaction) grows as 
$l_p$ decreases. The bimodality in the $n(x)$ distribution function suggests that the surface-induced 
toroid formation is a first order transition. The perturbative calculation reproduces qualitatively
all the features in the simulated density profiles.

We also considered the peeling and unbinding of adsorbed structures by applying force. These
results, which are of direct relevance to AFM experiments \cite{30}, show that the forced-unbinding 
transition is surprisingly highly cooperative. For all structures (racquet-like and toroids)
unbinding occurs over a narrow force range. The magnitude of the critical force $f_c$ for a fixed
value of $T$ and $l_s$ increases as $l_p$ decreases. From general considerations we expect that
$f_c$ should be described by a scaling a function $g(y)$ where $y$$=$$l_s/l_p$ for a fixed $T$. When 
$y$$<$$y_c$ (a critical value), then adsorbtion is not free energetically favored. When $y$$>$$y_c$, 
then $f_c$ should increase by an increasing function of $y$. The increase in $f_c$ can be achieved 
either by increasing $l_s$ for a fixed $l_p$ or by decreasing $l_p$ for fixed $l_s$. Additional
work is required to elucidate the nature of the scaling function $g(y)$. Quite surprisingly, we
find that the force-extension profiles can be calculated by using a simple perturbation theory
even though the nature of the unbinding transition is abrupt. The present work shows that
global properties of force-extension characteristics of adsorbed biomolecules can be nearly
quantitatively predicted using the proposed theory.

It is now well established that elastic response of DNA, in the absence of interaction with
surfaces, depends sensitively on the nature and concentration of counterions \cite{30,31}. Our
work shows that the force-extension curves in the presence of a surface to which DNA is bound
depends not only on $l_s$ but also on the morphology of the adsorbed structures. The novel
prediction that forced unbinding should occur cooperatively by a first-order phase transition
can be probed using single molecule experiments.

\acknowledgments{This work was supported in part by a grant from the National Science Foundation 
through NSF CHE05-14056. We would like to thank Changbong Hyeon for many useful discussions.}

\appendix
\section{Perturbative treatment of adsorbed chain statistics}

We expand ${\psi}_0$ (Eq. (\ref{1.1})) in terms of the Legendre polynomials, i.e.
\begin{equation}\label{2.3}
{\psi}_0 (x,\theta) = \sum_{i=0}^{\infty} {\psi}_i(x)P_i(\cos{[\theta]})  
\end{equation}
By using the following equations,
\begin{equation}\label{2.4}
{{d}\over {d\gamma}}P_i(\gamma) = -i(i+1)P_i(\gamma)\quad \text{and} \quad
P_1(\gamma)P_i(\gamma) = {{(i+1)P_{i+1}(\gamma)+iP_{i-1}(\gamma)}\over {2i+1}}
\end{equation}
we transform Eq. (\ref{1.3}) without term $\beta l_p \gamma f\psi_0$ into ($i\ge 0$)
\begin{equation}\label{2.5}
{{i(i+1)+\beta (u_{ads}-\epsilon_0)}\over {2l_p}}{\psi}_i(x) = -{{i}\over {2i-1}}
{{d{\psi}_{i-1}(x)}\over {dx}}-{{i+1}\over {2i+3}}{{d{\psi}_{i+1}(x)}\over {dx}}
\end{equation}
To the first few orders we have:
\begin{eqnarray}\label{2.6}
{\psi}_0 & = & {{2l_p}\over {3\beta (u_{ads}-\epsilon_0)}}{{d{\psi}_1}\over {dx}} \qquad (i=0),
\\\nonumber
{\psi}_1 & = & -{{2l_p}\over {2+\beta (u_{ads}-\epsilon_0)}}{{d{\psi}_0}\over {dx}}
-{{4l_p}\over {5(2+\beta (u_{ads}-\epsilon_0))}}{{d{\psi}_2}\over {dx}}\qquad (i=1),
\\\nonumber
{\psi}_2 & = & -{{4l_p}\over {3(6+\beta (u_{ads}-\epsilon_0))}}{{d{\psi}_1}\over {dx}}
-{{6l_p}\over {7(6+\beta (u_{ads}-\epsilon_0))}}{{d{\psi}_3}\over {dx}}\qquad (i=2),
\end{eqnarray}
${\psi}_1$ is given by the second Eq. (\ref{2.6}) with the second term determined by
${\psi}_2$ which is of the order of $8l_p^3d^3\psi_0/dx^3$. Neglecting this order for 
$\eta$$\ll$$1$ we obtain: 
\begin{equation}\label{2.7}
{\psi}_1 \approx -{{2l_p}\over {2+\beta (u_{ads}-\epsilon_0)}}{{d{\psi}_0}\over{dx}}
\end{equation}
Including the second term in the third Eq. (\ref{2.6}) and using Eq. (\ref{2.7}) we obtain 
the first-order perturbation equation for ${\psi}_1$,
\begin{equation}\label{2.8}
{\psi}_1 \approx -{{2l_p}\over {2+\phi}}{{d{\psi}_0}\over{dx}}
-{{4l_p}\over {5(2+\phi)}}{{d}\over {dx}}
\left[  {{4l_p}\over {3(6+\phi)}}{{d}\over {dx}} 
\left(  {{2l_p}\over {2+\phi}}{{d{\psi}_0}\over {dx}} \right) \right]
\end{equation}
where $\phi=\beta (u_{ads}-\epsilon_0)$. Substitute Eq. (\ref{2.8}) into the first Eq. 
(\ref{2.6}) we arrive at the first-order perturbative equation for ${\psi}_0$:
\begin{equation}\label{2.9}
{{64l_p^4}\over {45\phi(2+\phi)^2(6+\phi)}}{{d^4{\psi}_0}\over {dx^4}}+
{{4l_p^2}\over {3\phi(2+\phi)}}{{d^2{\psi}_0}\over{dx^2}} -{\psi}_0 = 0
\end{equation}
For the class of potentials considered here, the physical solution of Eq. (\ref{2.9}) that satisfies
the boundary conditions,
\begin{equation}\label{2.10}
\psi_0 (x=0) = 0, \quad \psi_0(x\to \infty)\to 0,
\quad {{d^n}\over {dx^n}} {\psi}_0 |_{x\to \infty} \to 0, \quad n=1,2,\ldots
\end{equation}
continuity requirements at $x=b$,
\begin{equation}\label{2.11}
\psi_0(x\to b-0) =\psi_0(x\to b+0),\quad 
{{d}\over {dx}}\psi_0 |_{x\to b-0} = {{d}\over {dx}}\psi_0 |_{x\to b+0}
\end{equation}
and normalization condition is given by Eq. (\ref{2.13}).

\section{Exact treatment of the chain distribution in the proximal range} 

Let us first consider the non-adsorbed chain in the presence of weak potential $u_{ads}$$\to$$0$. 
Assuming a self-similar distribution, ${\psi}_0(x,\Theta)$
$=$$x^{\alpha}$$g(\kappa)$ where $\kappa$$=$$\Theta(2l_p/x)^{1/3}$, we rewrite Eq. (\ref{3.1}) 
with $u_{ads}$$=$$\epsilon_0$$=$$0$ as an eigenvalue problem for $g(\kappa)$, 
\begin{equation}\label{3.3}
-{{1}\over {\kappa }}{{\partial^2 g}\over {\partial \kappa^2}}+
{{\kappa }\over {3}}{{\partial g}\over {\partial \kappa}} = \alpha g 
\end{equation}
Upon substitution $z$$=$$\kappa^3/9$, Eq. (\ref{3.3}) reduces to the following equation:
\begin{equation}\label{3.4}
z{{d^2 g}\over {dz^2}}+\left( {{2}\over {3}} - z \right){{dg}\over {dz}} +\alpha g = 0
\end{equation}
Under condition $g(z$$\to$$\infty)$$\to$$0$ the only solution to Eq. (\ref{3.4}) is 
$g(z)$$\sim$$\Psi(-\alpha,2/3,z)$ where $\Psi(\chi,\omega,z)$ and $\Gamma (\chi)$ are defined in
the main text. $\psi (x$$\to$$0,$$\Theta$$<$$0)$$\to$$0$ defines the spectrum of eigenvalues 
$\alpha_n$$=$$1/6$$-$$n$, where $n=0,$$\pm 1,$$\pm 2,$$\ldots$ (see Appendix B in Ref.~\cite{26}). 
The requirement that $\psi $ does not have knots is satisfied for $n=0$ ($\alpha$$=$$1/6$) and 
$n$$=$$1$ ($\alpha$$=$$-5/6$), and the general solution for $u_{ads}$$=$$0$ in the region $x$$>$$b$ 
is given by Eq. (\ref{3.7}) of the main text.

To solve Eqs. (\ref{3.8}) we substitute $C_0$ from the second equation into the first equation 
and multiply by $x^{2/3}$. We obtain:
\begin{equation}\label{A.1}
x^{4/3}{{d^2C_1}\over {dx^2}}-{{2}\over {3}}x^{1/3}{{dC_1}\over {dx}}+(\bar{F}_{11}+
Dx^{2/3})C_1=0
\end{equation}
where $D=\bar{F}_{00}\bar{F}_{11}-\bar{F}_{01}\bar{F}_{10}$ and $\bar{F}_{nm}=F_{nm}
\phi (2l_p)^{1/3}$, $n,m=0,1$. Substituting $y=x^{1/3}$ into Eq. (\ref{A.1}) and multiplying 
it by $y^2$, we get:
\begin{equation}\label{A.2}
y^2{{d^2C_1}\over {dy^2}}-2y{{dC_1}\over {dy}}+y^2(9\bar{F}_{11}+9Dy^2)C_1=0
\end{equation}
Using $z=y^2$ allows us to rewrite Eq. (\ref{A.2}) as
\begin{equation}\label{A.3}
\gamma_2 z{{d^2C_1}\over {dz^2}}+\beta_1{{dC_1}\over {dz}}+(\gamma_0 z+\beta_0)C_1=0
\end{equation}
where $\gamma_0$$=$$9D$, $\beta_0$$=$$9\bar{F}_{11}$, $\beta_1$$=$$-4$ and $\gamma_2$$=$$4$.
The general solution of Eq. (\ref{A.3}) is given by
\begin{equation}\label{A.5}
C_1(x) = e^{ {{3}\over {2}}\sqrt{-D} x^{{{2}\over {3}}} }
[ c_1 \Phi (\rho,-{{1}\over {2}}, -3\sqrt{-D} x^{{{2}\over {3}}} )
+ c_2(-3\sqrt{-D} x^{{{2}\over {3}}})^{{{3}\over {2}}} 
\Phi (\rho + {{3}\over {2}}, {{5}\over {2}}, -3\sqrt{-D} x^{{{2}\over {3}}})]
\end{equation}
where $c_1$, $c_2$ are constants and $\rho$$=$$(3$$\bar{F}_{11}$$-$$2\sqrt{-D})/$$4\sqrt{-D}$. 
In Eq. (\ref{A.5}) $\Phi (k,l,x)$ is the Kummer series defined in the text. In the range 
$0$$\le$$x$$\le$$b$, $\psi_1$$(x,\Theta$$=$$0)$ diverges as $x$$\to$$0$. To avoid this divergence 
we require that $C_1$$(x$$=$$0)$$=$$0$. This is satisfied when $c_1$$=$$0$. To insure that 
$C_1(x)$$\to$$0$ as $x$$\to$$\infty$ for $x$$>$$b$, we set $c_2$$=$$0$. Substituting 
$\phi$$=$$\phi_{in}$ and $\phi$$=$$\phi_{out}$ into solutions for $0$$\le$$x$$\le$$b$ and 
$x$$>$$b$ and using formulas $\Phi (k,l,x)$$=$$e^x$$\Phi (l-k,l,x)$, ${{d^m}\over {dx^m}}
\Phi (k,l,x)$$=$${{(k)_m}\over {(l)_m}}$$\Phi (k+m,l+m,x)$ we obtain Eqs. (\ref{3.10}).

\newpage

\section*{\bf FIGURE CAPTIONS}

{\bf Figure 1:}
Schematic of a semiflexible chain (blue) adsorbed on the surface (yellow). 
The free end $({\bf r},{\bf n})$ makes an angle $\theta$$=$$\arccos{[{\bf n}\cdot{\bf n}_x/|{\bf n}
|\cdot|{\bf n}_x|]}$ with the direction ${\bf n}_x$ of the pulling force ${\bf f}$$=$$f$${\bf n}_x$.
For clarity the chain is shown as extended which is realized only when the SC-surface interaction is
strong. The interaction between the monomers of the chain and the surface is attractive in the
range $0$$\le$$x$$\le$$b$ where $x$ is the distance perpendicular to the surface. The strength of
the interaction is $\Delta$. In the Langevin simulations we replace the square well potential by the
Lennard-Jones potential (Eq. (\ref{0.8})).

\bigskip

{\bf Figure 2:}
Top view of the typical structures (blue) adsorbed on the surface (yellow) 
for $S$$=$$120$$B$, $\Delta$$=$$1.5$$B$ and $k_B T$$=$$1.0$. Extended, one-racquet, two-racquet, 
three-racquet and toroidal structures are obtained in a single trajectory that is terminated at 
$t$$=$$4$$\times$$10^{-5}$$\tau$. The equilibrium structure under these conditions is the toroid.

\bigskip

{\bf Figure 3:}
Dependence of radius of gyration $R_g/a$ (top left), intramolecular attractive interaction $U_{LJ}$
(Eq. (\ref{0.6})), surface potential $U_{ads}$ (Eq. (\ref{0.8})), and the internal energy 
$U$ ($=$$U_{bend}$$+$$U_{bond}$$+$$U_{LJ}$) displayed as functions of time, measured in units of
$\tau$. The five curves in each panel correspond to extended (black), one-racquet (red), two-racquet 
(green), three-racquet (blue) and toroidal (magenta) structures of Fig. 2.

\bigskip

{\bf Figure 4:}
The average monomer density profiles $n(x)$ as a function of $x/b$ for $\Delta$$=$$1.5$$B$ and 
$k_B T$$=$$1.0$ for extended, two-racquet, three-racquet and toroidal states. The left panel is 
for $S$$=$$70$$B$ and the results for $S$$=$$50$$B$ are shown on the right. Solid lines and dotted
lines represent the results obtained using perturbative and exact theory, respectively.

\bigskip

{\bf Figure 5:}
Dynamics of extension $x$ (in units of $a$) for a few trajectories at different values of $f$
applied to the chain ends of structures shown in Fig. 2. Time $t$ is expressed in units of $\tau$.
The colors correspond to the caption in Fig. 3. The values of $f$ are displayed in the panels.

\bigskip

{\bf Figure 6:}
The averaged reduced extension $\langle x\rangle /L$ as a function of constant force $f$ simulated 
for $\Delta$$=$$1.5$$B$ and $k_B T$$=$$1.0$ for structures in Fig. 2 ($S$$=$$120$$B$, top). The
bottom panel shows force-extension profiles for four-, five-, seven-racquet and  toroidal 
configurations obtained for $S$$=$$30$$B$ (bottom panel). Data points for extended (four-racquet), 
one-racquet (five-racquet), three-racquet (seven-racquet) and toroidal structures are given by red, 
green, blue and black circles, respectively. Theoretical curves for these structures are given 
respectively by solid, dotted, dashed and dot-dashed lines.

\newpage

\begin{figure}
\includegraphics[width=7.00in]{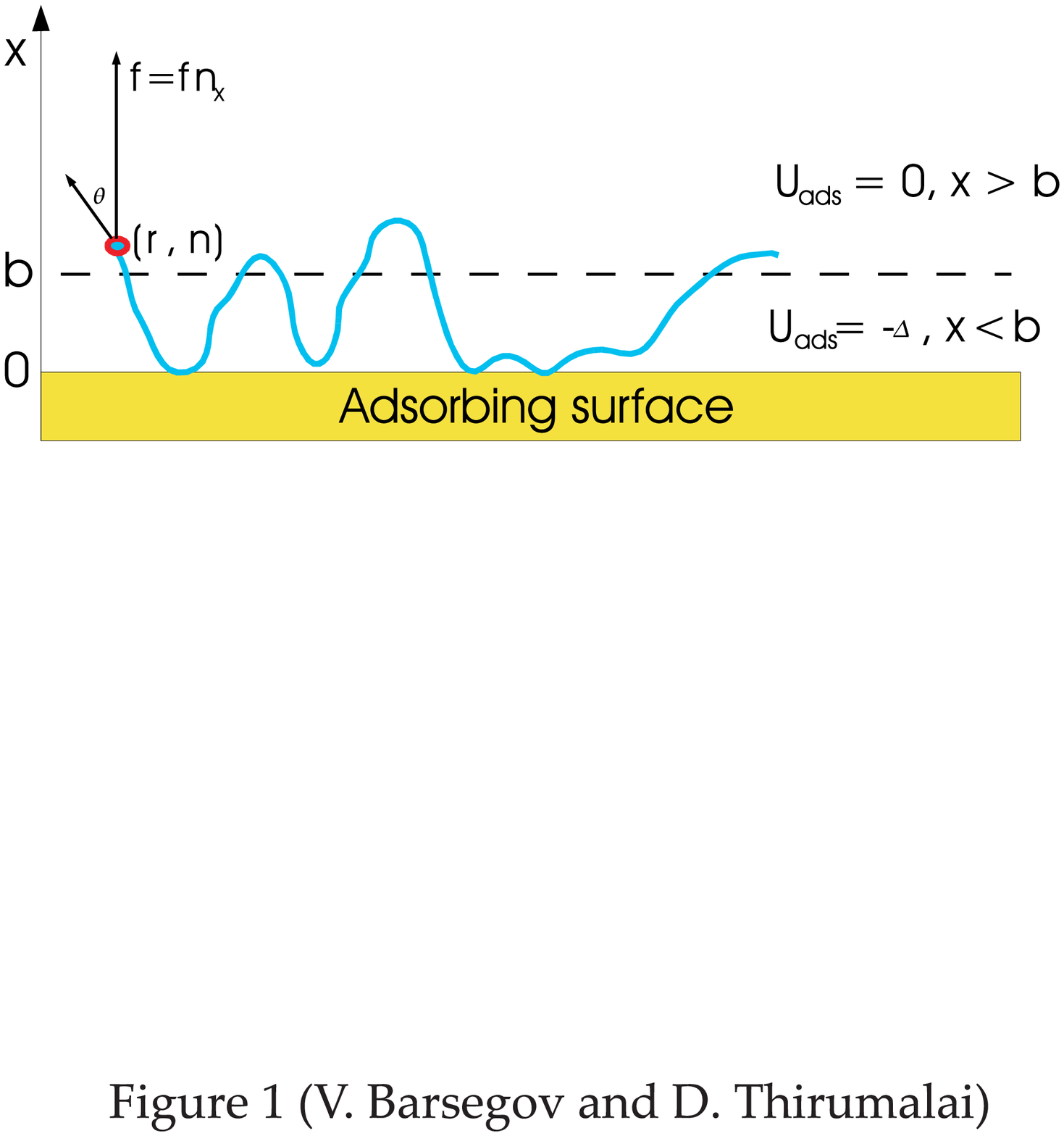}
\end{figure}

\newpage

\begin{figure}
\includegraphics[width=7.00in]{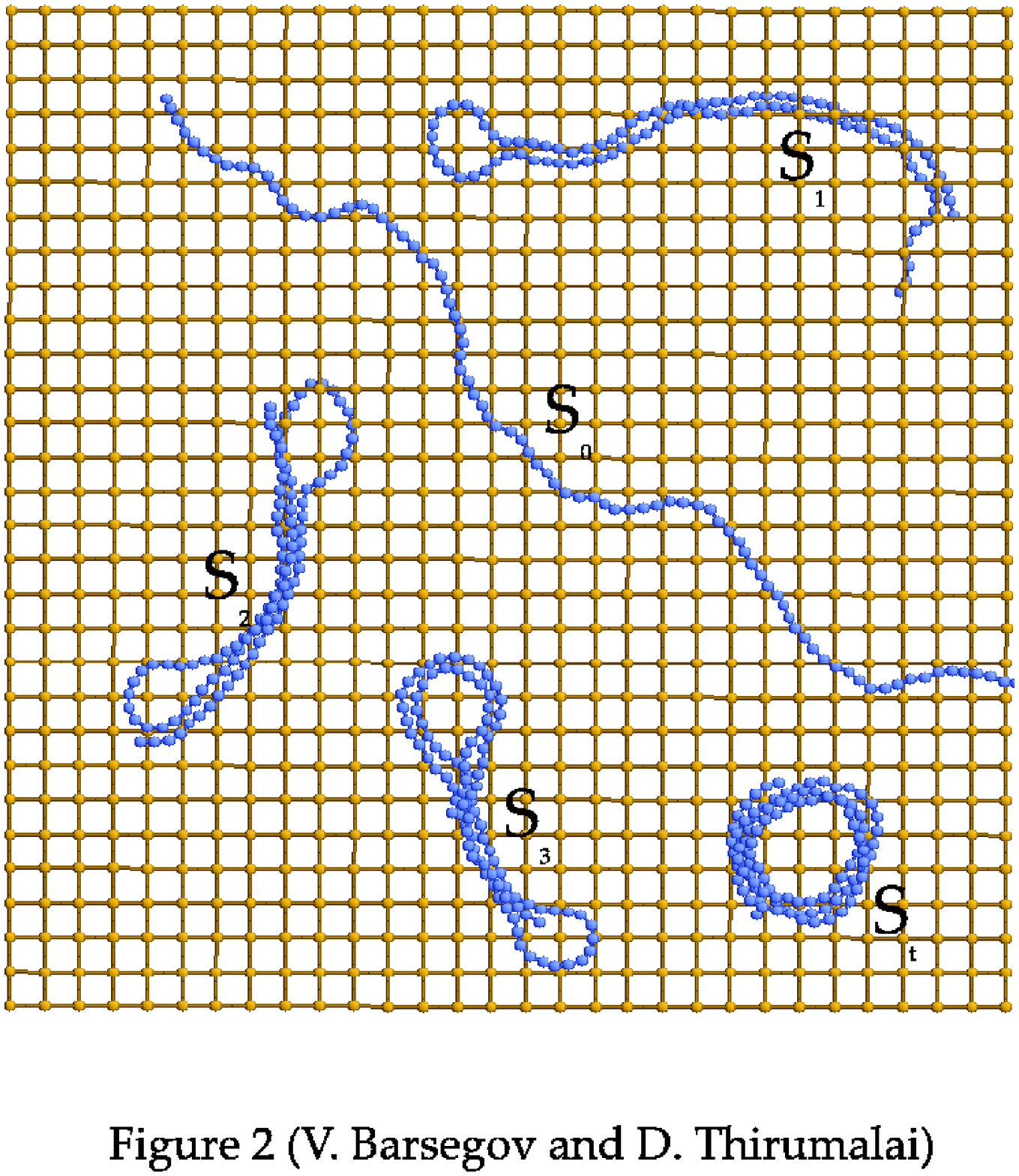}
\end{figure}

\newpage

\begin{figure}
\includegraphics[width=7.00in]{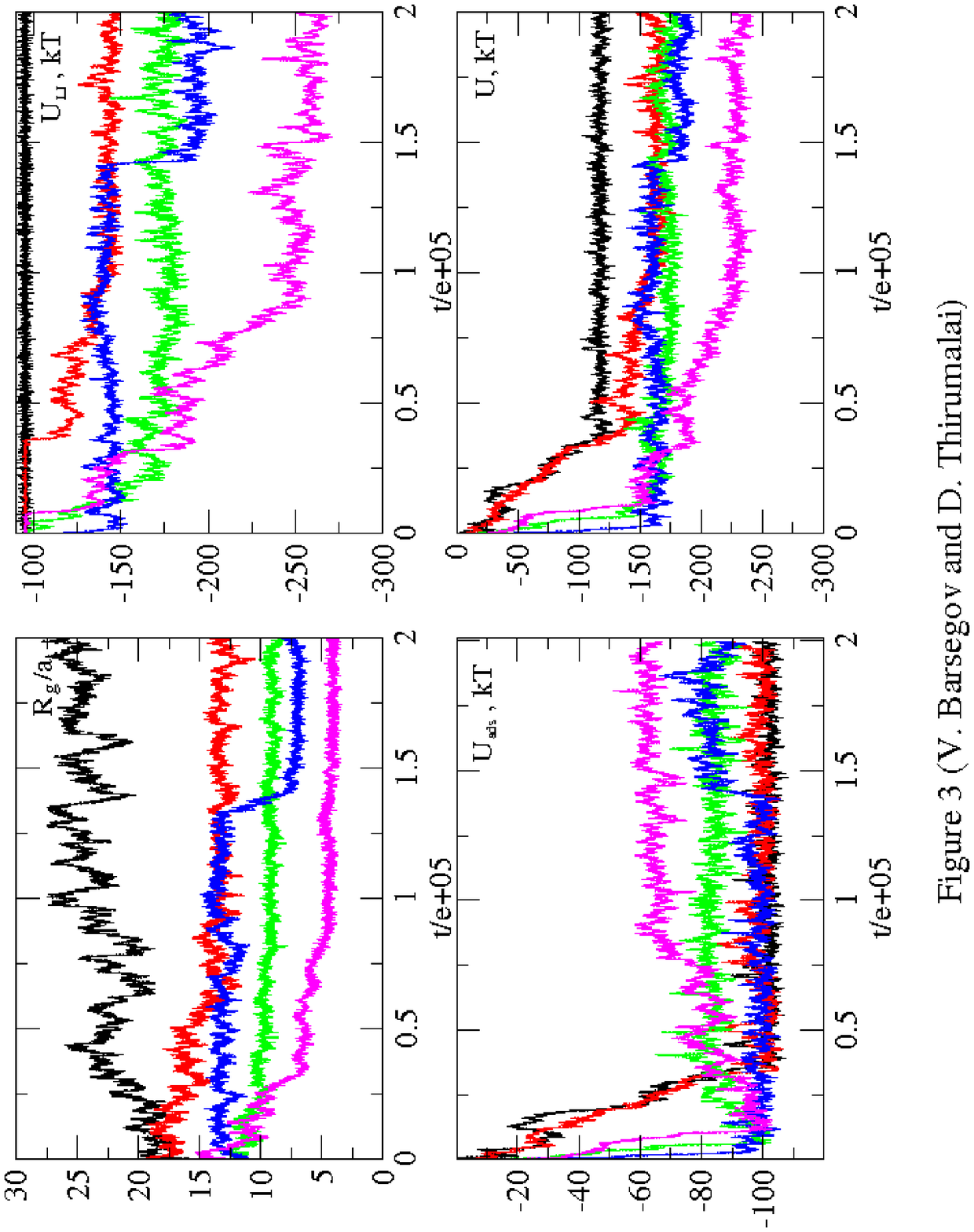}
\end{figure}

\newpage

\begin{figure}
\includegraphics[width=5.50in]{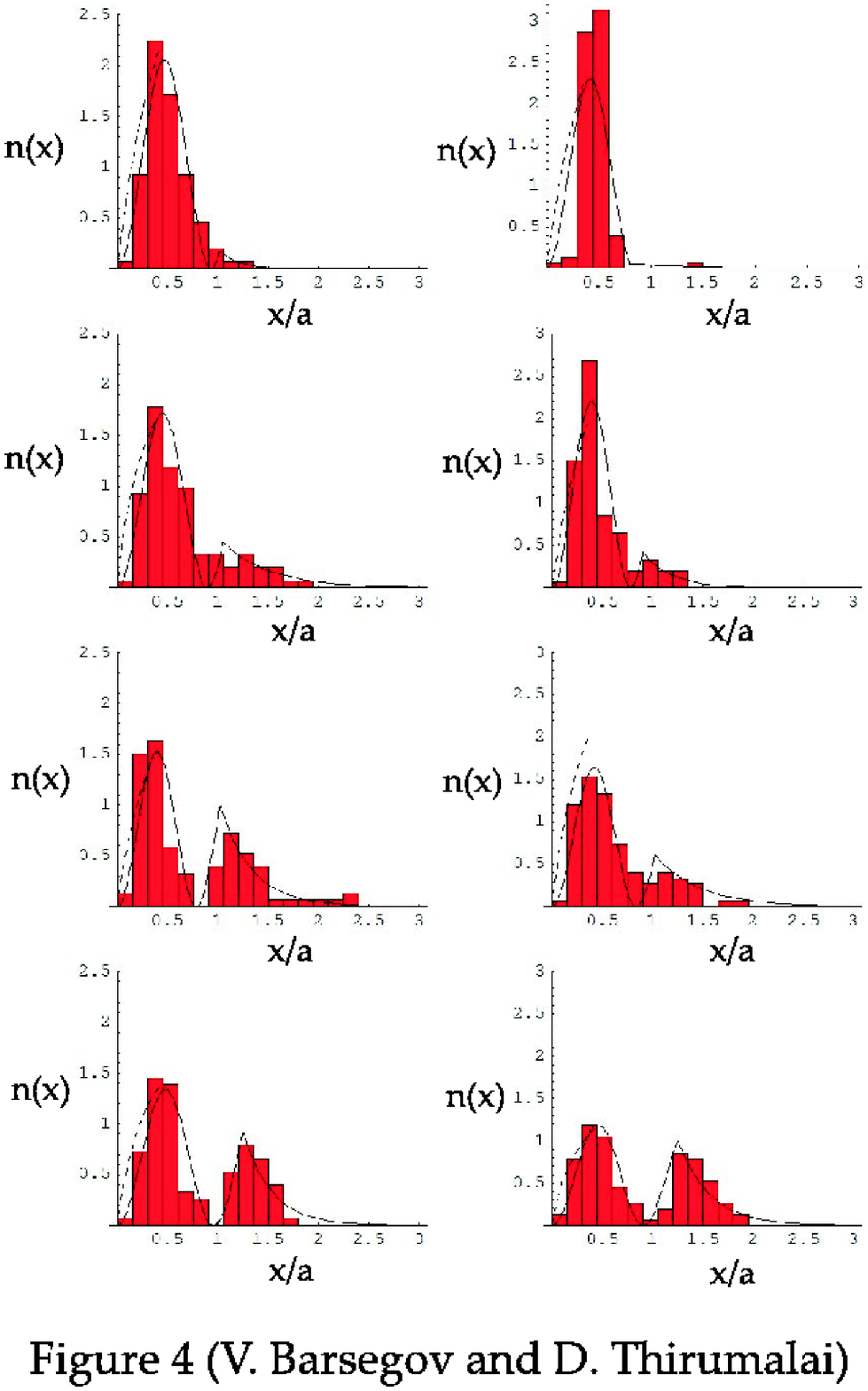}
\end{figure}

\newpage

\begin{figure}
\includegraphics[width=7.00in]{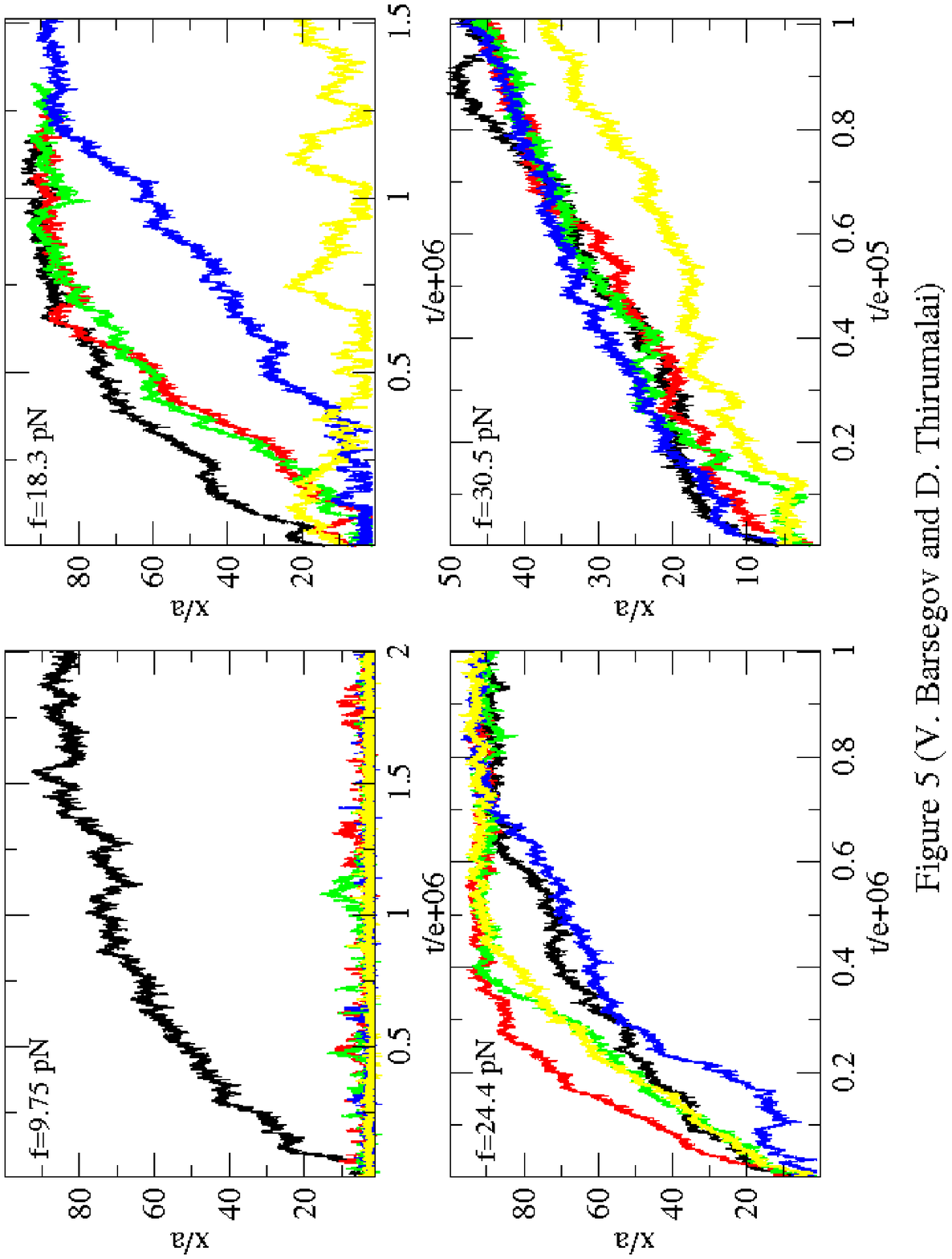}
\end{figure}

\newpage

\begin{figure}
\includegraphics[width=6.70in]{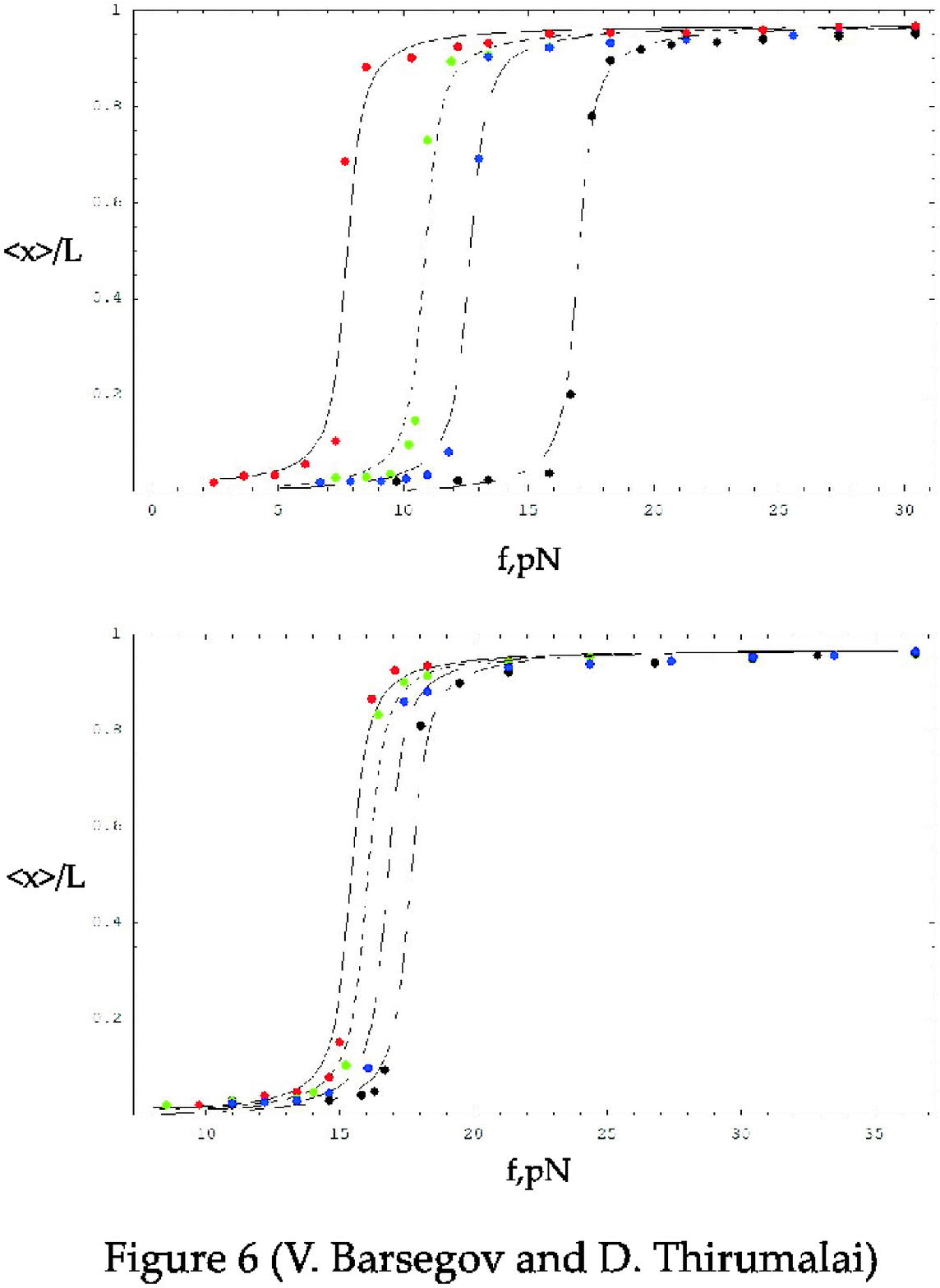}
\end{figure}

\end{document}